\documentclass[utf8,final]{frontiersSCNS}

\usepackage{url,hyperref,lineno,microtype,subcaption}
\usepackage[onehalfspacing]{setspace}

\def\keyFont{\fontsize{8}{11}\helveticabold }
\def\firstAuthorLast{Kowal \& Falceta-Gonçalves}
\def\Authors{Grzegorz Kowal\,$^{1,\dagger,*}$ and Diego A. Falceta-Gonçalves$^{1,\dagger}$}
\def\Address{$^{1}$ Escola de Artes, Ciências e Humanidades, Universidade de São Paulo, Av. Arlindo Béttio, 1000 -- Vila Guaraciaba, CEP: 03828-000, São Paulo - SP, Brazil}
\def\corrAuthor{Grzegorz Kowal}
\def\corrEmail{grzegorz.kowal@usp.br}
\def\authorContrib{$^{\dagger}$ These authors have contributed equally to this work and share first authorship.}

\begin{document}
\onecolumn
\firstpage{1}

\title[Colliding-Wind Binaries as a Source of TeV Cosmic Rays]{Colliding-Wind Binaries as a Source of TeV Cosmic Rays}

\author[\firstAuthorLast]{\Authors}
\address{}
\correspondance{}

\extraAuth{}

\maketitle

\authorContrib{}

\begin{abstract}
In addition to gamma-ray binaries which contain a compact object, high-energy
and very high-energy gamma rays have also been detected from colliding-wind
binaries. The collision of the winds produces two strong shock fronts, one for
each wind, both surrounding a shock region of compressed and heated plasma,
where particles are accelerated to very high energies. Magnetic field is also
amplified in the shocked region on which the acceleration of particles greatly
depends. In this work, we performed full three-dimensional magnetohydrodynamic
simulations of colliding winds coupled to a code that evolves the kinematics of
passive charged test particles subject to the plasma fluctuations. After the run
of a large ensemble of test particles with initial thermal distributions, we
show that such shocks produce a nonthermal population (nearly 1\% of total
particles) of few tens of GeVs up to few TeVs, depending on the initial
magnetization level of the stellar winds. We were able to determine the loci of
fastest acceleration, in the range of MeV/s to GeV/s, to be related to the
turbulent plasma with amplified magnetic field of the shock. These results show
that colliding-wind binaries are indeed able to produce a significant population
of high-energy particles, in relatively short timescales, compared to the
dynamical and diffusion timescales.

\tiny
 \keyFont{ \section{Keywords:} stars: Wolf–Rayet, stellar winds, gamma-ray sources, particle astrophysics, cosmic rays, turbulence, magnetohydrodynamics (MHD), methods: numerical}
\end{abstract}

\section{Introduction}

Star-forming regions are among the sources of high-energy (HE) photons in our
galaxy. The reason for that is still a subject of investigation, given that
these regions are populated by a mix of magnetized turbulent plasma, as well as
a number of stellar objects at different evolutionary stages, given their
initial masses, such as collapsing gas cores of stars yet to form, stars of main
sequence, as well as more evolved massive ones, which are already at the edge of
being subjects to energetic explosive events that end their evolution. Shocks
and magnetic fields, basic ingredients for particle acceleration, are present in
all the mentioned scenarios and give a hint of possible generation of HE photons
from HE particles by synchrotron, inverse Compton, and pion decay processes.
While the magnetized plasma of the interstellar medium is understood as
important for the particle diffusion and scattering, stellar sources and compact
objects have been mentioned as main drivers of particle acceleration in
star-forming regions (SFRs).

Supernova (SNe) remnants have been well associated as sources of HE photons and
particles for many decades. Among these, the type-II, Ib, and Ic SNe are
strongly related to the loci of star formation, given their progenitor's high
mass. Accretion –- and associated disks and jets -– is also commonly related to
the observed HE photons from SFRs, for example, the X-ray binaries. While
low-mass X-ray binaries (LMXB) are spatially distributed over the whole galaxy,
the high-mass counterpart (HMXB) is associated to SFRs \cite[see,
e.g.,][]{Grimm_etal:2002}.

Not only massive binaries with compact objects emit HE photons, strong shocks
generated by high-velocity winds of both stars in massive binary systems can
also accelerate particles to very high energies, being observed in hard X-rays
and gamma-rays bands, the so-called colliding-wind binaries (CWBs). These
objects are also strongly related to SFRs. Particle acceleration to high
energies from CWBs has been suggested by \citet{EichlerUsov:1993}. Indeed,
nonthermal emission -- especially in radio frequencies -- associated to high
energy particles interacting with local magnetic fields is often detected from
these objects \citep[][]{DeBeckerRaucq:2013}.

The theory of diffusive shock acceleration (DSA) is based on the scattering and
change of particle momenta as they interact with plasma waves in a second order
Fermi acceleration fashion \citep[e.g.,][]{Fermi:1949, Bell:1978}. As pointed by
\citet{EichlerUsov:1993}, there is the need for a relatively strong magnetic
field at the shock region for the particles to be efficiently accelerated. The
minimum strength of magnetic field can be estimated in order for the
collisionless regime to occur; that is, the gyrofrequency must be larger than
the thermal collision frequency. There is a number of recent works attempting to
estimate HE particle populations and the related emission of radiation for CWBs
that do not properly consider the distribution of magnetic fields in the shocks
\citep[e.g.,][]{Dougherty_etal:2003, Reimer_etal:2006, Reitberger_etal:2014a,
Reitberger_etal:2014b, delPalacio_etal:2016, Pittard_etal:2020}.

These previous works, although very useful to generally describe spectral
features of nonthermal and high-energy sources associated to CWBs, have not been
able to provide a proper answer to two major open issues in this field: 1) how
the particles are accelerated in CWBs -- which includes how fast this
acceleration occurs and how they diffuse, and 2) where these particles are
mainly accelerated. To answer these two questions, a more complex analysis is
needed, based on full magnetohydrodynamics (MHD) combined with a full treatment
of particle acceleration either by evolution of momentum distributions or, even
better, by studying the kinematics of individual particles.

Full MHD simulations were first employed for the study of nonthermal emission
from CWBs by \citet{FalcetaGoncalvesAbraham:2012}, which showed that parallel
shocks are unrealistic, and magnetic field amplification at the downstream is
much larger than expected from adiabatic modeling. Full MHD simulations of CWBs
were further presented by \citet{Kissmann_etal:2016}. Geometry complexity and
amplification are products of strong cooling of the post-shocked plasma, which
brings a scenario much different from the ones typically assumed for
semi-analytical approximations. In \citet{FalcetaGoncalvesAbraham:2012},
however, the particle acceleration was treated in a simplified model, and the
energy distribution of particles could not be properly determined.
\citet{FalcetaGoncalves:2015} studied the first- and second-order Fermi
acceleration processes finding that second-order effects are dominant for their
MHD numerical simulations of eta Carinae. Energies as high as $1 - 10$~TeV, at
timescales of $10^4 - 10^5$~s, were given as parametric estimates for that
particular MHD setup of eta Carinae, although the energy distribution of
particles was not determined.

Given the context described above, a natural improvement of the theoretical
modeling of particle acceleration in CWBs would be the development of a full
kinetic description of particles as subject to magnetized plasma fluctuations
derived from a full MHD simulation of the shocks. This is the main objective of
the present work. Here, we provide the first model to integrate a population of
thermal hadronic particles, subject to the local distributions of plasma
velocity, density, and magnetic fields derived from a full MHD numerical
simulation. The organization of the manuscript is given as follows.

In \S\ref{sec:modeling}, we describe the magnetohydrodynamic (MHD) model of the
CWB system we simulated and the method to evolve acceleration and diffusion of
test particles in this system. In \S\ref{sec:results}, we describe the results
of the MHD model and the most important characteristics of nonthermal particles
accelerated in the studied system. We also discuss the spatial distribution of
the acceleration rate. In \S\ref{sec:conclusions}, we briefly discuss our
results and draw our main conclusion.

\section{Numerical Modeling}
\label{sec:modeling}

\subsection{Magnetohydrodynamic Model of Colliding-Wind Binary}
\label{sec:simulation}

We model a binary system consisting of primary and secondary stars placed at a
distance of approximately 7.2~AU. The stars are described numerically by
spherical regions of the radius of about 0.12~AU at which the density, pressure,
outflow velocity, and magnetic field are maintained during the whole simulation,
and determined by emitted stellar winds. The wind density of the primary star is
set to $\rho_{m}=10^7$ in the code units (c.u.), while in the secondary star, it
is $10^2$ times smaller. We assumed the wind speeds of the primary and secondary
stars to be $500 \, \mathrm{km/s}$ and $2500 \, \mathrm{km/s}$, respectively.
The wind pressure at the star surfaces is determined by the formula $p_{wind} =
\left( v_{wind} / {M}_{wind} \right)^2 \rho_{wind} / \Gamma$, where ${M}_{wind}$
is the sonic Mach number of the wind at the stellar surface and $\Gamma$ is the
adiabatic index set to $1.1$. In our model, we set ${M}_{wind}$ to 3 and 15 for
the primary and secondary stars, respectively. Although magnetized, the massive
stellar winds are super-Alfv\'enic, that is, the magnetic pressure is smaller
than the kinetic one. As a consequence, near the star, the magnetic field lines
follow the velocity field profile, which is radial. The radially oriented fields
follow a $r^{-2}$ scaling, anchored at the stellar surface foot points. The foot
points are constant in time and are determined considering an internal magnetic
field of $10^{-3}$, in the code units, oriented along the $Z$ direction.  The
surface and internal magnetic fields are kept constant during the simulation,
while the plasma outside of the stars is evolved according to the MHD equations.
For the sake of setup simplicity, we also set initially the magnetic field of
the interstellar gas as uniform, with the same intensity. This, however, is
irrelevant for the results  presented here as the whole plasma of the
interstellar medium is pushed out of the computational domain as the winds
expand and reach the quasi-steady-state condition. It is important to notice
that the interstellar plasma parameters analyzed further in this work are
therefore independent of this external initial setup, depending only on the
stellar surface wind parameters.

The above setup is evolved by the ideal adiabatic compressible MHD system of
equations:
\begin{equation}
\frac{\partial \rho}{\partial t} + \nabla \cdot \left( \rho \mathbf{u} \right) = 0,
\label{eq:mhd_1}
\end{equation}
\begin{equation}
\frac{\partial \left( \rho \mathbf{u} \right)}{\partial t} + \nabla \cdot \left[ \rho \mathbf{u} \mathbf{u} + \left( p + \frac{B^2}{2} \right) I - \mathbf{B} \mathbf{B} \right] = 0,
\label{eq:mhd_2}
\end{equation}
\begin{equation}
\frac{\partial \mathbf{B}}{\partial t} - \nabla \times \left( \mathbf{u} \times \mathbf{B} \right) = 0, \, \nabla \cdot \mathbf{B} = 0,
\label{eq:mhd_3}
\end{equation}
\begin{equation}
\frac{\partial E}{\partial t} + \nabla \cdot \left[ \left( E + p \right) \mathbf{u} - \left( \mathbf{u} \cdot \mathbf{B} \right) \mathbf{B} \right] = 0,
\label{eq:mhd_4}
\end{equation}
where $\rho$, $p$, $\mathbf{u}$, and $\mathbf{B}$ are the plasma density, thermal
pressure, velocity, and magnetic field, respectively, $E = (\rho u^2/2) +
p/(\Gamma - 1) + (B^2/2)$ is the total energy, and $I$ is the identity matrix in a
3D rectangular domain described on the Cartesian coordinate system with physical
dimensions $9L \times 6L \times 6L$ (assuming $L \approx 1.2$~AU) corresponding
to the X, Y, and Z directions, respectively, the base resolution of $48 \times
32 \times 32$, and the local mesh refinement up to the $5^{th}$ level resulting in
the effective resolution of $768 \times 512 \times 512$ (the effective grid size
being $h \approx 0.012 \, \mathrm{c.u.} \approx 0.014 \, \mathrm{AU}$ along all
directions). The refinement criterion was based on the normalized value of
vorticity and current density to follow the shock, turbulence, and magnetic
reconnection development. We also set the outflow boundaries along all
directions.

The MHD equations were solved numerically using a high-order shock-capturing
adaptive refinement Godunov-type code {\tt AMUN}\footnote{AMUN code is open
source and freely available from
\href{https://gitlab.com/gkowal/amun-code}{https://gitlab.com/gkowal/amun-code}.
} with 5$^{th}$-order optimized compact monotonicity preserving reconstruction
of Riemann states \citep{AhnLee:2020}, the HLLD Riemann flux solver
\citep{Mignone:2007}, and the 3$^{rd}$ order 4-step Strong Stability Preserving
Runge-Kutta (SSPRK) method \citep{Gottlieb_etal:2011} for time integration. The
system was evolved until $t = 21 \, \mathrm{c.u.}$ (corresponding to roughly 44
days considering the assumed length and velocity units), at which the
interaction surface between the colliding winds stopped moving through the
domain, and a turbulent region produced by their interaction became
statistically stationary. Once such a state is reached, the system can be
evolved for an arbitrary period of time maintaining its statistical properties.

\subsection{Test Particle Integration}
\label{sec:particles}

In order to study the behavior of test particles in the binary system evolved as
described in the previous subsection, we took the last available domain snapshot
(velocity and magnetic field spatial distributions) at $t = 21$ and injected
$10^5$ (if not specified otherwise) protons in random locations near the
turbulent region created by colliding winds with random orientations of velocity
and the initial particle velocity distribution corresponding to the thermal
distribution at temperature $T = 10^6 \, \mathrm{K}$.

The trajectories of injected particles were integrated using the relativistic
equation of motion for a charged particle
\begin{equation}
  \frac{d}{d t} \left( \gamma \mathbf{v} \right) = \frac{q}{m} \left( \mathbf{E} + \mathbf{v} \times \mathbf{B} \right), \quad \frac{d \mathbf{x}}{dt} = \mathbf{v}, \label{eq:motion}
\end{equation}
where $\mathbf{x}$ and $\mathbf{v}$ are the particle position and velocity,
respectively, $\gamma = \left( 1 - v^2/c^2 \right)^{1/2}$ is the Lorentz factor,
$q$ and $m$ are particle charge and mass, respectively, $c$ is the speed of
light, and $\mathbf{E} = - \mathbf{u} \times \mathbf{B}$ is the plasma-induced
electric field, with $\mathbf{u}$ and $\mathbf{B}$ being the plasma velocity and
magnetic field taken from the grid simulation, respectively.

We assumed the particle velocity $\mathbf{v}$ in Equation~(\ref{eq:motion}) to
be expressed in units of the speed of light $c$ and its position in the
coordinates of the grid simulation. Since the grid simulations use dimensionless
units, one needs to assume the unit of plasma velocity $\mathbf{u}$ and magnetic
field $\mathbf{B}$. In this work, we assumed that the plasma velocity is in
units of $1000$~km/s, and for the magnetization level of the winds, we set the
magnetic field anchored at the stellar surfaces to range from $B_{\star} =
10^{-5}$ to $10^{0}$~G. Both stars are assumed to maintain the same intensity of
surface magnetic fields during the whole simulation time. This value is a model
parameter and should not be confused with the real stellar magnetic strength.
Due to the numerical modeling limitations, it simply represents the magnetic
field which is ejected from the star together with the wind. We should also note
that in the integration, the unit of time was chosen as $50 \, \mathrm{h} = 1.8
\times 10^5$~s which, assuming the plasma velocity unit of $1000$~km/s, results
in the unit of length equal to $L \approx 1.2 \, \mathrm{AU}$.

The integration of Equation~(\ref{eq:motion}) was done with the help of the code
{\tt GAccel}\footnote{GAccel code is open source and freely available from
\href{https://gitlab.com/gkowal/gaccel}{https://gitlab.com/gkowal/gaccel}.}
using the 8$^{th}$ order embedded Dormand-Prince (DOP853) method
\citep{Hairer_etal:2008} with an adaptive time step based on an error estimator.
In order to determine the values of $\mathbf{u}$ and $\mathbf{B}$ at the current
particle positions, we used tri-cubic interpolation (a cubic interpolation
applied dimension by dimension), enforcing the monotonicity by limiting the
tangents at the interpolation nodes with the minmod TVD limiter
\citep{vanLeer:1979}.  In the integration of the particle trajectories, we set
the absolute and relative tolerance to {\tt atol = rtol = 1e-10} and the initial
time step {\tt dt = 1e-6}, if not specified otherwise.

\section{Results}
\label{sec:results}

\subsection{Characteristics of the Colliding-Wind System}

The model of colliding-wind binary described in \S~\ref{sec:simulation} was
evolved up to the equilibrium position of the contact discontinuity of the shock
apex ($t \approx 21$~c.u.). At this time, the region at which the winds interact
reach stochastic saturation. It is characterized by the presence of strong
turbulence produced by the interacting winds and magnetohydrodynamic
instabilities. This region is also characterized by strong compression, allowing
for the increase of the magnetic field energy to the levels above the magnetic
energy maintained within the stars.

\begin{figure}[t]
\begin{center}
\includegraphics[width=0.95\columnwidth]{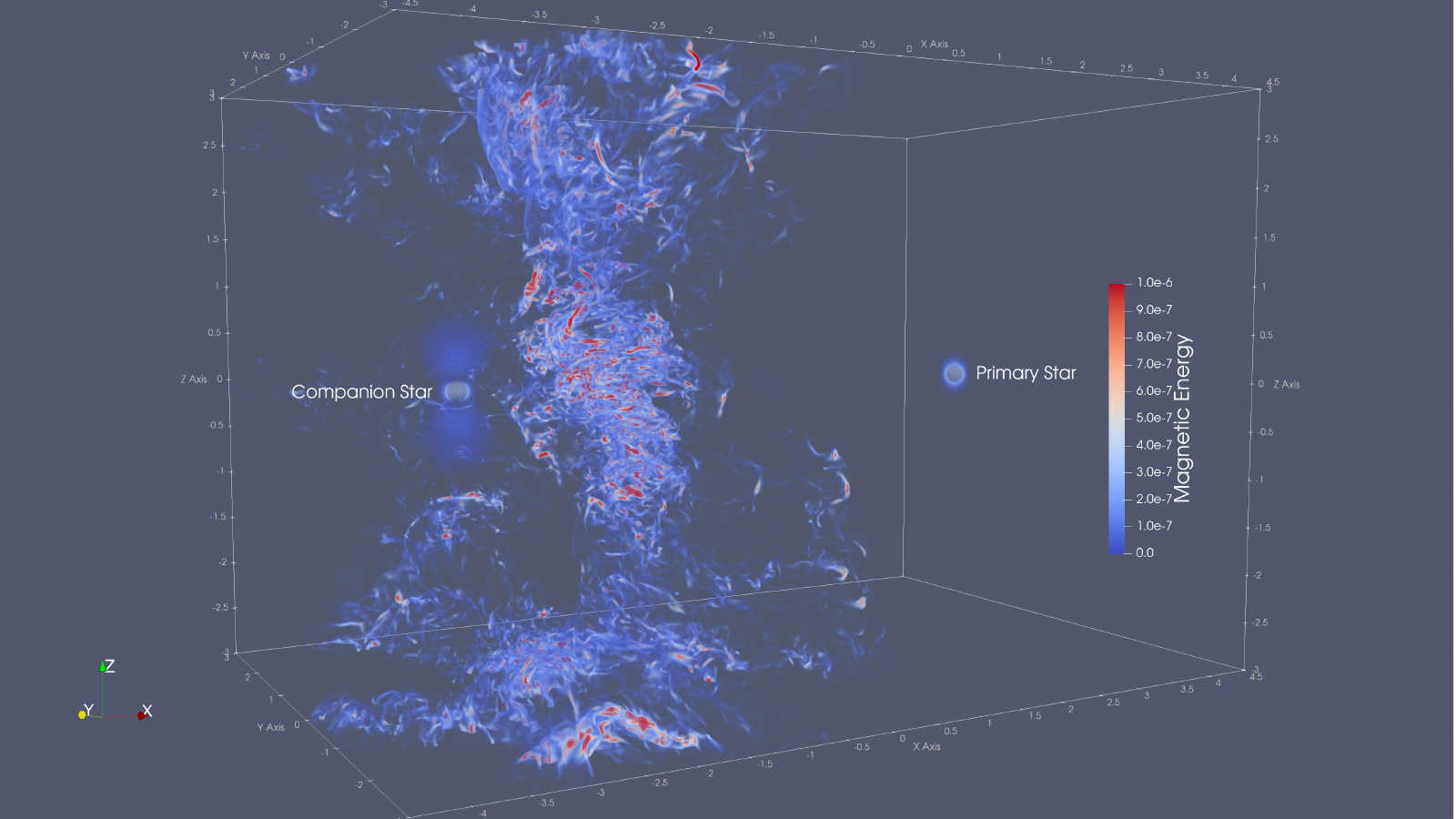}
\end{center}
\caption{Distribution of magnetic energy in the computational domain of the
studied colliding-wind binary system at time $t = 21$. On the right, we indicate
the primary star, and on the left, the secondary one. The turbulent region
produced by the interacting winds is characterized by a number of clumps of
increased magnetic energy. Regions of magnetic energy larger than the energy
within the stars are shown in red.}
\label{fig:Eb-distribution}
\end{figure}

\begin{figure}[t]
\begin{center}
\includegraphics[width=0.95\columnwidth]{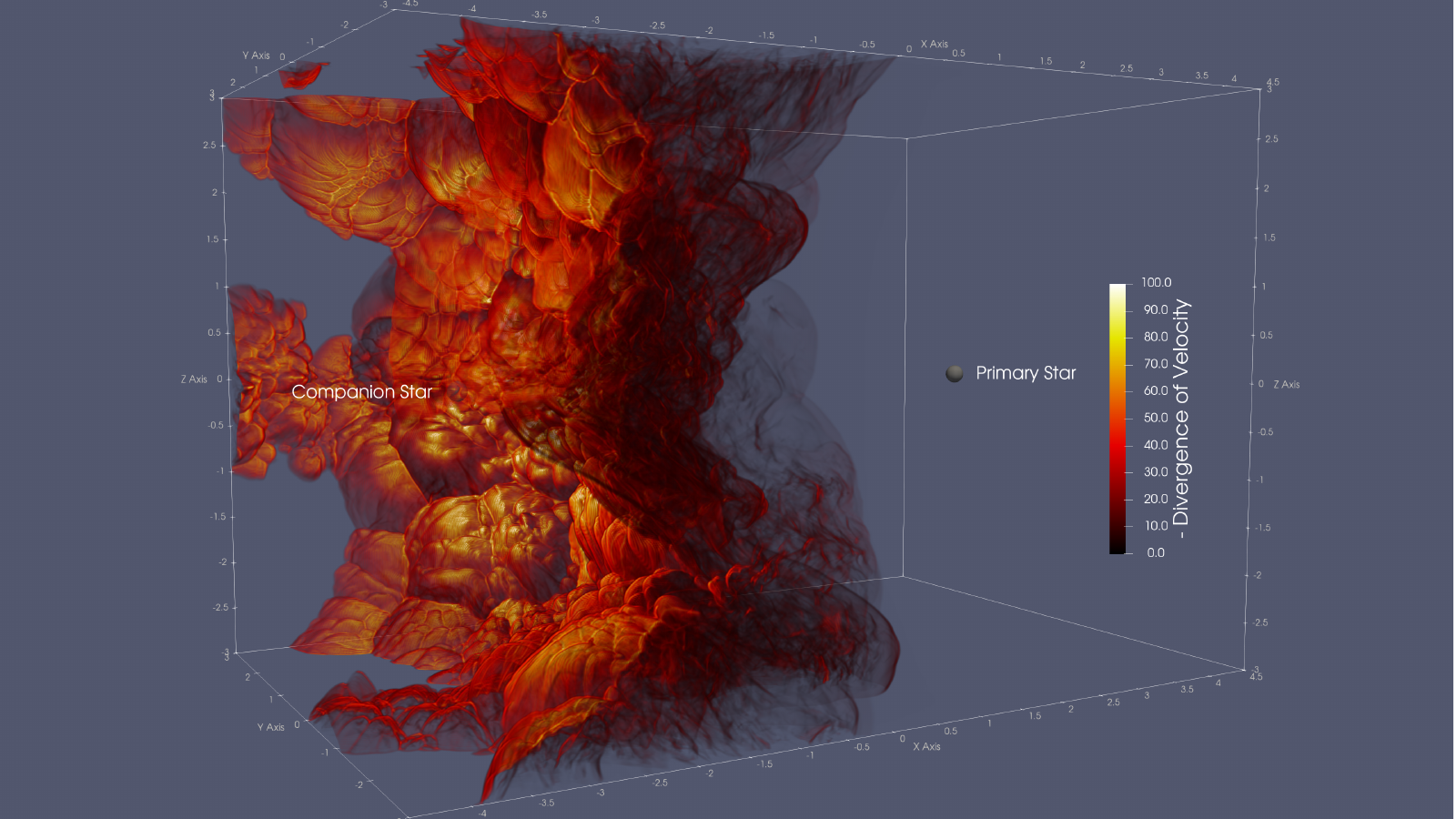}
\end{center}
\caption{Distribution of the divergence of velocity in the computational domain
at time $t=21$ in the same projection as Fig.~\ref{fig:Eb-distribution}. Only
the positive values of $-\nabla\cdot\mathbf{u}$ are shown. The volume where the
gas is affected by the wind interaction can be clearly seen. It is characterized
by the development of strong compressing shocks and turbulence.}
\label{fig:divV-distribution}
\end{figure}

In Figure~\ref{fig:Eb-distribution}, we show a 3D visualization of the magnetic
energy distribution at time $t=21$. The magnetic energy maintained in both stars
is equal to $5\times10^{-7}$c.u. (white, in the middle of the presented color
bar). However, $E_{mag}$ quickly decays as the magnetic field is transferred by
the wind from the stellar surfaces and distributed into the surroundings (see
blueish envelopes around the stars). Since the winds are super-Alfvéenic, the
field decays radially, that is, $r^{-2}$. It is again compressed in the region
of the wind interaction forming strongly magnetized structures (reddish clumps
in Fig.~\ref{fig:Eb-distribution}). The maximum magnetic field strength in these
structures could reach values in an order of magnitude higher than those set at
the stellar surfaces. The distribution of these magnetic clumps is clearly
related to the turbulent region produced by the interacting winds.

In Figure~\ref{fig:divV-distribution}, we show the distribution of the
divergence of velocity ($\nabla \cdot \mathbf{u}$) in the same projection as the
previous figure. For clarity, the values have inverted sign, so, for example,
the value $100$ indicates $\nabla \cdot \mathbf{u} = -100$. It is evident that
the negative part of the divergence of plasma velocity is a good tracer for
determining the extension of the volume affected by the interaction of the
colliding winds. Expressing the largest values seen in
Figure~\ref{fig:divV-distribution} in the units of $u = 1000 \, \mathrm{km/s}$
and $L \approx 1.2 \, \mathrm{AU}$ gives the maximum compression rate of around
$3.46 \, \mathrm{h}^{-1}$, indicating high compression efficiency of these
interactions. The regions of compressed gas relax the increased pressure
converting it into the mechanical energy and producing turbulence and pushing
away the turbulence structure. This compression-relaxation process can be
maintained in such a system as long as the stars emit their powerful winds.

In order to allow for comparison of observations with our results on the
particle acceleration and maximum energies reached, which are presented in the
following sections, we have estimated the mean magnetic field $\bar{B}_{shock}$
within the shock region of the performed numerical modeling of CWBs. The region
was determined by the condition $\nabla \cdot \mathbf{u} \le -50$. We have
obtained the mean intensity of the magnetic field in the shock region to scale
as $\bar{B}_{shock} \sim 10^{-1} B_{\star}$, where, as mentioned earlier,
$B_{\star}$ is the model parameter determining the strength of the magnetic
field at the distance of 0.12~AU from the star center. The variance of the
magnetic field strength in the shock region was estimated to be of the same
order as the mean value.

\subsection{Cosmic Rays Produced by the Colliding-Wind System}

As we explained in \S~\ref{sec:particles}, the protons are injected within the
turbulent region of the interacting winds. More precisely, we injected $10^5$
protons into the region $\left[ -1.5, -0.5 \right] \times \left[ -1.0, 1.0
\right] \times \left[ -1.0, 1.0 \right]$ of the computational domain
corresponding to region limits along the $x$, $y$, and $z$ coordinates,
respectively. The protons are integrated up to $t_{max} = 5 \times 10^4 \,
\mathrm{hours}$ or until they reach the boundary of the domain leaving the
system, whichever occurs first.

As an example, we provide a movie demonstrating the evolution of protons during
the time of integration for the model with $B_{\star} = 10^{-1}$ in the
Supplementary Material. In the movie protons are represented as color spheres,
with the color corresponding to the kinetic energy level $E_k$. The integration
time elapses in the logarithmic scale.

\begin{figure}[t]
\begin{center}
\includegraphics[width=\columnwidth]{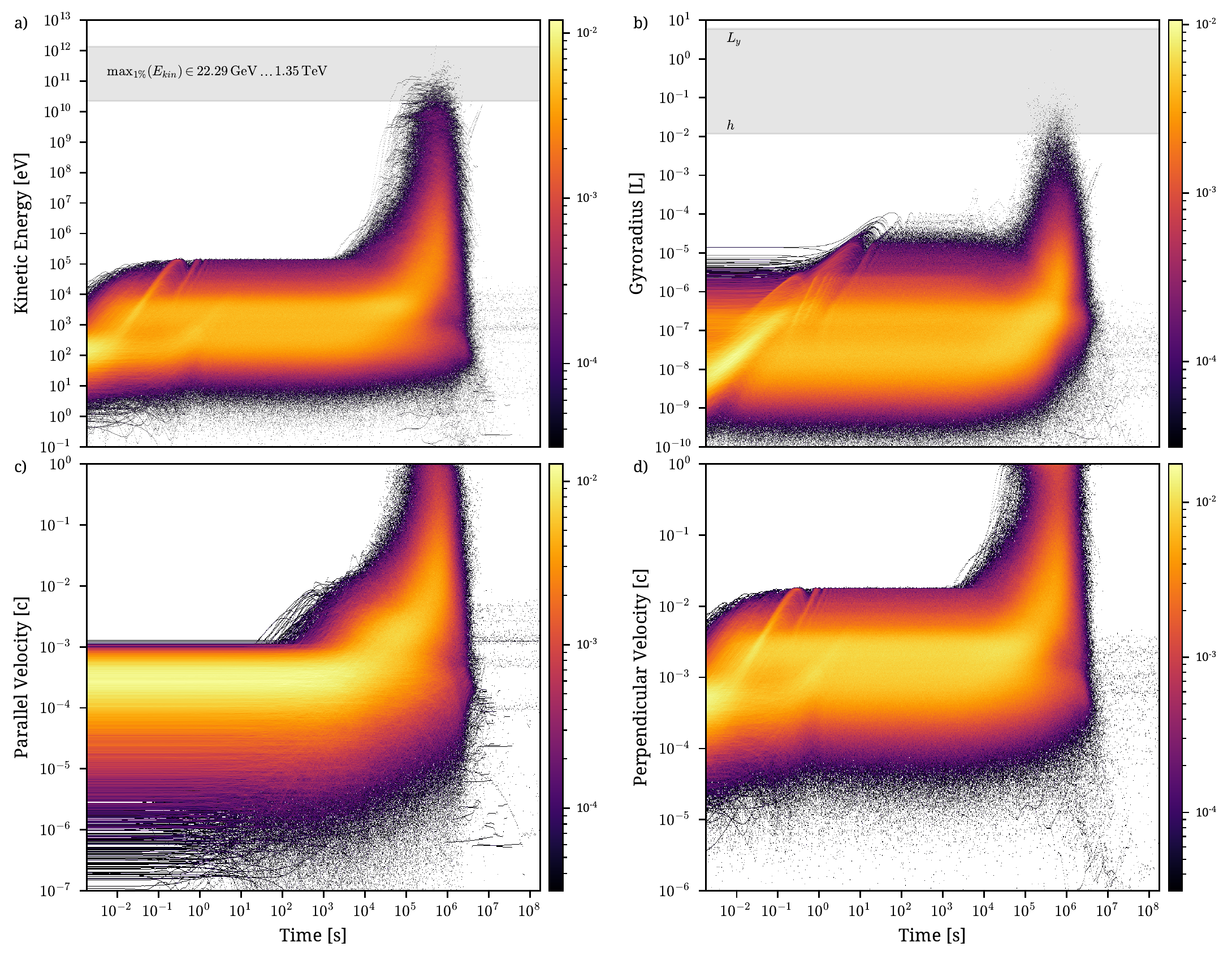}
\end{center}
\caption{Evolution of distributions of the particle kinetic energy (\textbf{a}),
the gyroradius (\textbf{b}), and the parallel and perpendicular particle
velocities (\textbf{c}) and (\textbf{d}) for the model with $B_{\star} = 10^{-1}
\, \mathrm{G}$. In the energy plot, the gray area indicates the range of the
maximum kinetic energy reached by 1\% of the most energetic protons. In the
gyroradius plot, the gray area indicates the scales permitted by the MHD
simulation, with $h$ being the effective cell size $h=1/512$ and $L_y = 6 L
\approx 7.2~\mathrm{AU}$ the size of the domain in the Y and Z directions.}
\label{fig:evolution}
\end{figure}

In Figure~\ref{fig:evolution}, we show the evolution of distributions of a few
most important properties characterizing the protons: (a) the evolution of
particle kinetic energy distribution, (b) the evolution of particle gyroradius
in the code units, (c) and (d) the evolution of parallel and perpendicular, with
respect to the local magnetic field, components of velocity in the units of $c$.
Figure~\ref{fig:evolution}a demonstrates that the protons suffer initially a
very quick, within a few seconds, pre-acceleration, most probably due to the
electromagnetic plasma drift or magnetic field curvature. At this stage, they
can reach the kinetic energies up to a few hundreds of MeV. In our analysis,
this value was insensitive to the value of $B_{\star}$ used in the integration.
The distribution of gyroradia does not change significantly, although some sort
of a reorganization is seen characterized by a population of protons for which
the gyroradius increases (see the period of time before $t = 100 \, \mathrm{s}$
in Fig.~\ref{fig:evolution}b). The increase of gyroradia seems to be related to
the increase of perpendicular velocities, since similar patterns are observed in
Figure~\ref{fig:evolution}d during this period. Contrary to the perpendicular
component, the parallel component of particle velocity is essentially unaltered.

The situation significantly changes at around $t = 200\,\mathrm{s}$ when the
parallel component starts to increase, initially to around one percent of $c$
between $t = 10^3 \, \mathrm{s}$ and $10^4 \, \mathrm{s}$, and eventually
reaching nearly the speed of light between $t = 10^5 \, \mathrm{s}$ and $10^6 \,
\mathrm{s}$ (see Fig.~\ref{fig:evolution}c). At this stage, that is, after $t =
10^5 \, \mathrm{s}$, one can notice a significant increase of the perpendicular
velocity component, reaching the speed of light, as well. Clearly, the increase
of the perpendicular velocity results in the increase of the gyroradius,
observed in Figure~\ref{fig:evolution}b. It is interesting to see, however, that
even due to relativistic particle speed, only a small fraction of protons has
gyroradia larger than the grid size $h$, barely reaching the specified length
unit of $L \approx 1.2 \, \mathrm{AU}$.

As seen in Figure~\ref{fig:evolution}a, after $t = 10^4 \, \mathrm{s}$, a
significant fraction of protons can increase their kinetic energies, defined as
$E_{k} \equiv \left( \gamma - 1 \right) m_p c^2$, where $m_{p}$ is the proton
mass, by several orders of magnitude, from levels below $E_{k} \sim 0.1 \,
\mathrm{MeV}$ to tens and hundreds of GeV. The gray area seen in this plot
indicates the range of kinetic energies reached by 1\% of the most energetic
protons. We should clarify that this range does not represent the energies at
which protons left the computational domain but the maximum energy they reached
during the whole integration. As we explained above, the energetic level after
the pre-acceleration at $t < 10^4 \, \mathrm{s}$ is essentially insensitive to
the used strength of the magnetic field $B_{\star}$. As we will see below, the
energy band of the most energetic protons, on the contrary, is strongly
sensitive to $B_{\star}$.

\begin{figure}[t]
\begin{center}
\includegraphics[width=0.95\columnwidth]{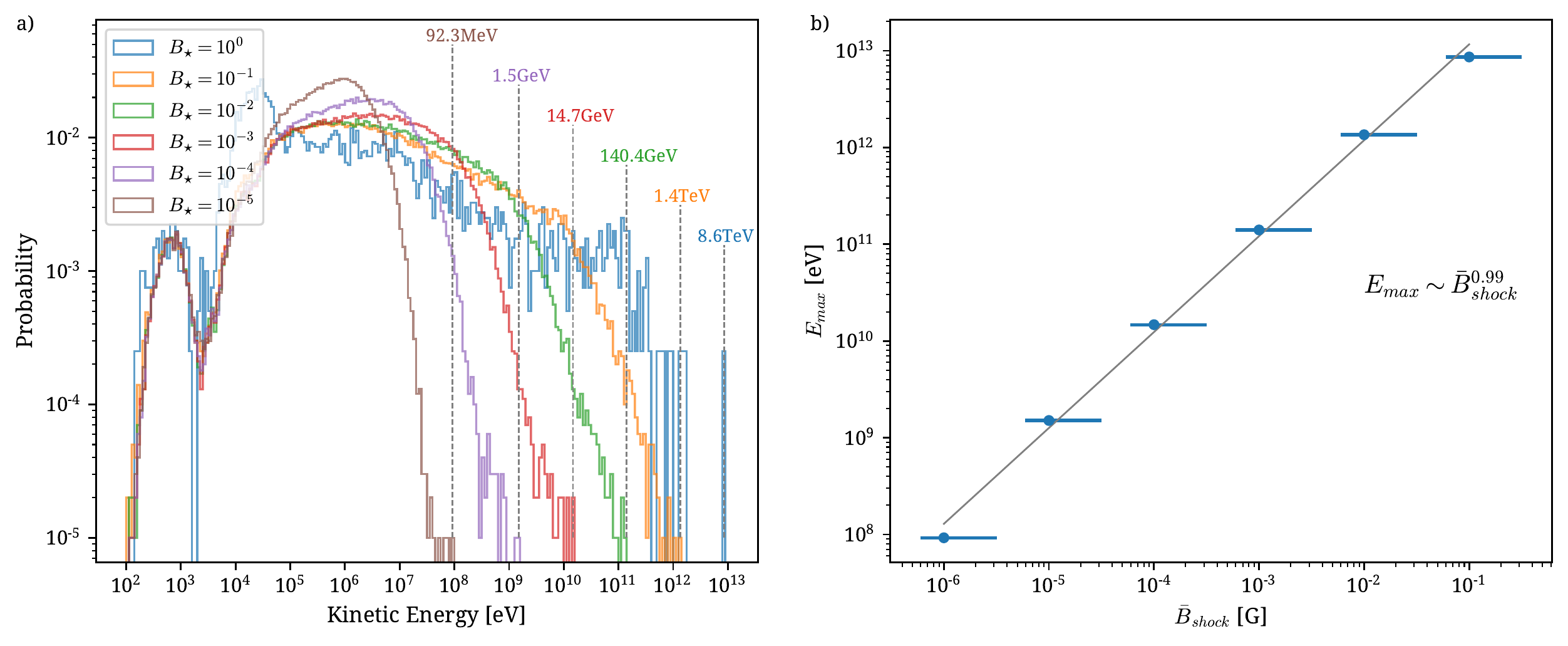}
\end{center}
\caption{{\em Left}: Distribution of the particle largest kinetic energy for
integration done with different values of $B_{\star}$, from $10^{-5}$ to $10^{0}
\, \mathrm{G}$. Vertical gray-dashed lines indicate the maximum kinetic energies
reached in each integration. {\em Right}: Dependence of the maximum kinetic
energy (indicated by the vertical lines in the left panel) on the mean magnetic
field in the shock region $\bar{B}_{shock}$.}
\label{fig:emax}
\end{figure}

In the left plot of Figure~\ref{fig:emax}, we show distributions of the maximum
energies reached by protons for all studied particle integration cases, that is,
for $B_{\star} = 10^{-5} \, \mathrm{G}$ to $1 \, \mathrm{G}$. As one can see,
the low-energy population $E_{k} < 10^4 \, \mathrm{eV}$ is very similar for all
cases. Also, the second more energetic population of protons is relatively
insensitive to the magnetic field strength. Nevertheless, from around the energy
level of $10^6 \, \mathrm{eV}$ up, one can identify that the high-energy tail of
these distribution strongly depends on $B_{\star}$. The vertical dashed lines
indicate the maximum energy reached in each case. The relation between these
energies and the values of $\bar{B}_{shock}$, the mean field in the shock
region, with its dispersion as horizontal error bars, is shown in the right plot
of Figure~\ref{fig:emax}. It is interesting to see that the maximum energy
scales nearly linearly with the magnetic field strength. Moreover, from this
plot, we can see that around 0.1\% or more of very energetic protons in the
range of 100~GeV to 1~TeV can be produced for $\bar{B}_{shock} = 10^{-3} \,
\mathrm{G}$ or stronger. This dependence also allows to extrapolate the maximum
energies for colliding-wind systems characterized by even stronger magnetic
fields, indicating that at least one particle of the energy over 10~TeV could be
produced once every around 4 months (most protons typically leave the system
within $10^7 \, \mathrm{s}$, as seen in Fig.~\ref{fig:evolution}).

\begin{figure}[t]
\begin{center}
\includegraphics[width=0.95\columnwidth]{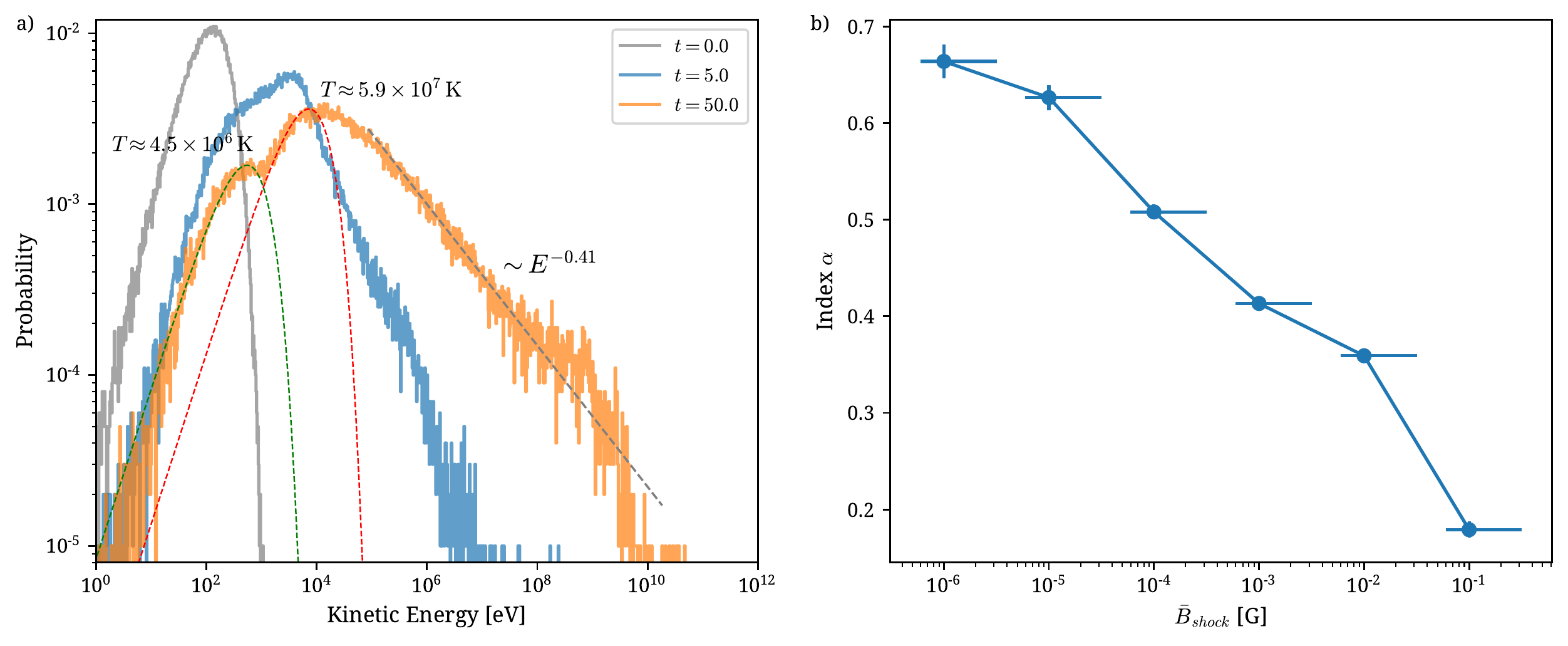}
\end{center}
\caption{{\em Left}: Particle kinetic energy distribution at moments $t=0$,
$t=5$, and $t=50$ hours (gray, blue, and orange, respectively) for the
integration done with $B_{\star} = 10^{-2} \, \mathrm{G}$. At the initial time,
the distribution is purely thermal with temperature $T = 10^6 \, \mathrm{K}$. At
the final time shown, the distribution is characterized by two thermal
populations with $T \approx 4.5 \times 10^6 \, \mathrm{K}$ (green dashed line)
and $T \approx 5.9 \times 10^7 \, \mathrm{K}$ (red dashed line) and the
high-energy power-law tail (gray dashed line) with the fitted index of around
$-0.4$. {\em Right}: The dependence of the power-law index fitted to the
nonthermal tail on mean magnetic field in the shock region $\bar{B}_{shock}$ at
the same moment $t = 50$.}
\label{fig:energy-histogram}
\end{figure}

The maximum energies shown here were obtained neglecting any process of energy
loss. We are basically interested in the acceleration of hadrons, that is, the
typical electron energy losses such as synchrotron emission, inverse Compton,
Coulomb collisions, and bremsstrahlung emission losses are unimportant and may
be disregarded. The protons are expected to suffer energy loss mainly due to
hadronic collisions. The process of pion generation, and further decay, is the
main energy loss phenomenon at the energy range obtained from our simulations.
In order to validate our results, we may estimate the loss rate for proton-proton
collisions, as given by \cite{Kelner_etal:2006} and \cite{Pittard_etal:2020}:
\begin{equation}
\frac{d \gamma}{dt} \simeq \frac{1}{2} n_p \gamma_p c \sigma_{pp},
\label{eq:pploss}
\end{equation}
with $\sigma_{pp}$ being
\begin{equation}
\sigma_{pp} = (34.3 + 1.88 \lambda + 0.25 \lambda^2)\left[ 1 - \left( \frac{1.22 \, {\rm GeV}}{E_{p}}\right)^2\right]^2 \times (10^{-27} {\rm cm}^2),
\end{equation}
where $\lambda = \ln(E_{p}/1{\rm TeV})$, $E_{p} \equiv \gamma m_{p} c^2$ is the
particle total energy, and $n_p$ is the proton number density. Equation
\ref{eq:pploss} was solved for each time step of the particle integration
procedure. The deceleration rates $d \gamma / dt$ for the nonthermal energetic
particles result in maximum energy loss scaling as $\sim 10^{-10} n_p
\left(\bar{B}_{shock} / \mathrm{1 G} \right)$ GeV/s, where $\bar{B}_{shock}$ is
the mean magnetic intensity in the shock region. The dependence on the thermal
proton number density $n_p$ has been kept in order to allow us to estimate under
what conditions the energy loss rate would be relevant compared to the
acceleration process.

As discussed in the next subsection, the acceleration rates for these particles
lie in the range of few to several GeV/s. In order to halt the acceleration
process, the hadronic collisions will be dominant, in a mG shocked region, only
for proton number densities above $10^{13}$ cm$^{-3}$, which is well above any
typical post-shock condition expected for the massive systems known. Therefore,
the estimates of acceleration and energy distributions obtained here are likely
to be independent on these loss terms.

The analysis of the maximum energies of protons produced in the colliding-wind
system gives an insight into the possible potential of the system in the
production of high or very high energetic protons. In order to analyze how the
thermal protons reach higher energies, we have to analyze the particle
distributions at given times. In the left plot of
Figure~\ref{fig:energy-histogram}, we demonstrate the distributions of the
kinetic energy for the integration done with $B_{\star} = 10^{-2} \, \mathrm{G}$
for three different moments: the initial thermal distribution of $T = 10^6 \,
\mathrm{K}$ (gray line), at the intermediate stage at $t = 5 \, \mathrm{h}$ when
the high-energy power-law tail starts to develop, and the distribution at $t =
50 \, \mathrm{h}$, just before the protons start leaving the computational
domain. We can notice that the purely thermal protons are initially heated
toward slightly higher temperatures and develop a nonthermal energetic tail. At
$t = 50.0$, the distribution is characterized by two thermal populations of
protons (fitted by green and red dashed curves) with temperatures estimated to
around $4.5 \times 10^6 \, \mathrm{K}$ and $5.9 \times 10^7 \, \mathrm{K}$ and a
nonthermal power-law tail, $dN \propto E^\alpha \, dE$, with the index estimated
to be around $\alpha \approx -0.41$ (gray dashed line). In the right plot of
Figure~\ref{fig:energy-histogram}, we show the dependence of the fitted
power-law index on mean magnetic field strength of the turbulent region,
$\bar{B}_{shock}$. The index depends on the value of $\bar{B}_{shock}$ decaying
from around 0.66 for $\bar{B}_{shock} = 10^{-6} \, \mathrm{G}$ to around 0.18
for $\bar{B}_{shock} = 0.1 \, \mathrm{G}$. The flattening of the power-law tail
is well justified, considering that in all cases the initial thermal
distribution has the same temperature and the energy band of the tail as well as
the energy gain efficiency increase with the value of $\bar{B}_{shock}$.

Observations of nonthermal emission from WR-O binary systems have been very
successful in the radio wavelengths, typically associated to the synchrotron
emission from relativistic electrons. The nonthermal leptonic emission at radio
wavelengths presents spectral indices as steep as $\alpha \sim -0.62$ with a few
systems, where $\alpha$ is in the range from $-0.4$ to $-0.2$
\citep{DoughertyWilliams:2000}. Radio emission observations of the galactic
super star cluster Westerlund 1 revealed some nonthermal radio sources with the
spectral index $\alpha$ in the range from $-0.6$ to $-0.06$
\citep{Dougherty_etal:2010}. Since we integrated hadron trajectories in our
simulations, we cannot overinterpret our results for electrons and related
synchrotron emission spectra.

Emission from relativistic hadrons, however, such as nonthermal bremsstrahlung
and $\gamma$-rays from pion decay, has only been systematically observed for the
system of $\eta$ Carinae. {\it Fermi-LAT} detected very energetic emission $>10$
GeV close to the periastron passage \citep{Abdo_etal:2010}. Emission from
$\eta$Car, in the range of 100 GeV up to $\sim 400$ GeV, has also been detected
by {\it H.E.S.S.} \citep{Abdalla_etal:2020}. The combined data in the range from
few up to hundreds of GeVs show a spectral index of $-0.7$ to $-0.5$. Since the
proton-proton interactions generate $\gamma$-ray spectrum with a slope similar
to that of the energy distribution of the parent protons, we may compare these
observation estimates to our results shown in
Figure~\ref{fig:energy-histogram}b. It suggests that by varying the typical
magnetic field intensity of the stellar surface $B_{\star}$, we can also obtain
the spectral indices in a compatible range from $-0.2$ to $-0.7$. Clearly, we do
not take into account the particle energy losses for obtaining the particle
energy spectra, but still the fact that we could obtain similar spectral indices
is encouraging and suggests future studies taking into account the radiative
energy losses and transfer, or more complex magnetic field stellar geometries.

\subsection{Acceleration Rate Distribution}

Besides the basic properties of the acceleration mechanism such as the maximum
kinetic energy reached by the protons, their distribution, and the index of the
nonthermal tail, it is also interesting to determine what the range of the
acceleration rates available in the given system is and the spatial distribution
of the sites where the efficient acceleration is favorable.

\begin{figure}[t]
\begin{center}
\includegraphics[width=0.95\columnwidth]{f6}
\end{center}
\caption{Distribution of the maximum acceleration rate $dE_{k}/dt$ in the
computational domain for the integration done with $B_{\star} =
10^{-1}~\mathrm{G}$. Regions indicated by red color correspond to sites where
the detected maximum energy gain rate was above 1~TeV/h. The maximum
acceleration rate detected in this model was nearly 34~TeV/h.}
\label{fig:acc-rate}
\end{figure}

In Figure~\ref{fig:acc-rate}, we show the spatial distribution of such
acceleration sites in our computational domain for the integration of protons
done with $B_{\star} = 10^{-1} \, \mathrm{G}$. There is a clear correlation of
these sites with the turbulent region produced by the interacting winds. Protons
crossing the sites shown with the red color could gain the energies at rates as
large as 1~TeV/h, with the maximum acceleration rate determined for this case
reaching nearly $34 \, \mathrm{TeV/h}$ ($\sim 9.3 \, \mathrm{GeV/s}$).

\begin{figure}[t]
\begin{center}
\includegraphics[width=0.6\columnwidth]{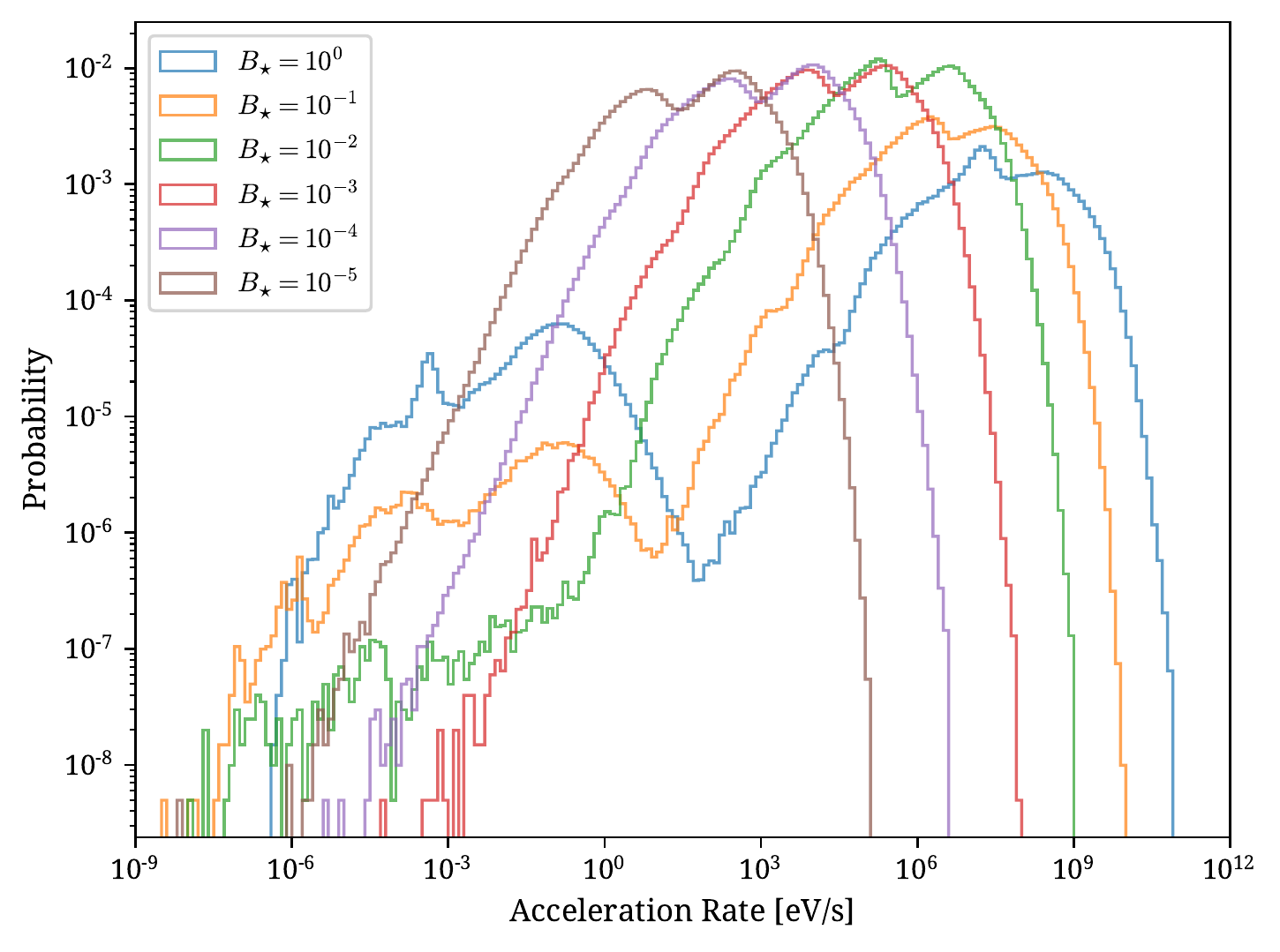}
\end{center}
\caption{Probability distribution function of the maximum acceleration rate
$dE_{k}/dt$ for the integration done with $B_{\star}$ from $10^{-5}$ to
$10^{0}~\mathrm{G}$. Relatively high fraction of the computational volume is
characterized by higher values of the acceleration rates, from keVs to GeVs per
second. The maximum acceleration rate is as high as nearly $80 \,
\mathrm{GeV/s}$ for the model with $B_{\star} = 1 \, \mathrm{G}$.}
\label{fig:acc-rate-dist}
\end{figure}

In Figure~\ref{fig:acc-rate-dist}, we show the distributions of the acceleration
rate for all studied particle models. There are a few important characteristics
seen in these distributions. First of all, the maximums are shifted toward
higher values of the acceleration rates for stronger magnetic fields. Moreover,
the bands where the most common rates are located depend in the magnetic field
strength $B_{\star}$. With the increasing $B_{\star}$, the acceleration rates
become more efficient, reaching orders of hundreds of GeV per second for the
most magnetized case $B_{\star} = 1 \, \mathrm{G}$. We should note that in this
case, we were able to integrate only 4,000 protons, which explains the visible
deviation of this distribution from other cases.

\section{Discussion and Conclusion}
\label{sec:conclusions}

In this work, we have investigated a colliding-wind binary system as a possible
site of high-energy and very high-energy cosmic ray production. We have
performed the magnetohydrodynamic simulations of interacting high velocity and
high mass loss rate stellar winds from a typical massive binary system. The
winds are characterized by supersonic speeds, $2500 \, \mathrm{km/s}$ and $500
\, \mathrm{km/s}$, for the primary and secondary stars, respectively. Collision
of such strong magnetized winds causes the production of shocks, turbulence, and
magnetic reconnection. These constitute fundamental ingredients for efficient
diffusive acceleration. The magnetic reconnection has been proposed as an
alternative acceleration mechanism, also represented by the first-order process,
gaining more attention recently \cite[see,
e.g.,][]{deGouveiaDalPinoLazarian:2005, LyubarskyLiverts:2008, Kowal_etal:2011,
Kowal_etal:2012, deGouveiadalPinoKowal:2015, delValle_etal:2016,
ComissoSironi:2019}.

In order to study the energy gain of nonthermal protons in such system, we
injected test particles and integrated their trajectories using the relativistic
equation of the charged particle motion. We present the evolution of
distributions of the particle kinetic energy, gyroradia, and parallel and
perpendicular, with respect to the local magnetic field, components of velocity.
In order to get a deeper insight in the possible energies, the protons can reach
in such a system where we estimate how the maximum kinetic energy and nonthermal
power-law index depend on the average magnetic field strength in the shock
region $\bar{B}_{shock}$. Finally, we determine the spatial distribution of the
maximum acceleration rate and its typical and maximum values.

Based on our results, we draw the following conclusions:

\begin{itemize}
\item The colliding-wind binary studied in this manuscript is able to produce
strongly magnetized turbulent region with shock being able to compress the
magnetic field to values higher than the magnetic field maintained within the
stars. For the limited resolution used in our modeling, the maximum magnetic
field strength in the system is about an order of magnitude higher than the
magnetic strength, $B_{\star}$, at a distance 0.12~AU from the star. Such a
turbulent region surrounded by incoming plasma favors the production of
nonthermal protons.

\item The highest available energies achieved by the accelerated protons depend
on the strength of magnetic field. For $\bar{B}_{shock} \ge 10^{-2} \,
\mathrm{G}$, 1\% of most energetic protons, a significant amount, reached
energies between tens and hundreds of GeV. For stronger magnetic field, the
level between hundreds of GeV and a few TeV was observed. We estimated that the
relation between the maximum energy particle and magnetic field scales nearly as
$E_{max} \propto \bar{B}_{shock}$. This result confirms that acceleration of
thermal protons up to the range of 1-10~TeV is feasible in the colliding-wind
binary systems.

\item For all studied values of the magnetic field intensity, we observe the
production of nonthermal tail in the energy distribution with the index
decreasing from around 0.7 for the weakest $\bar{B}_{shock} = 10^{-6} \,
\mathrm{G}$ to around 0.2 for $\bar{B}_{shock} = 0.1 \, \mathrm{G}$.

\item The regions of the highest acceleration rates $dE_{k}/dt$ are essentially
limited to the turbulent region produced by the colliding winds. In the case of
the strongest magnetic field studied here, $B_{\star} = 1 \, \mathrm{G}$, the
PDF of $dE_{k}/dt$ shows the more common range between a few tens of MeV/s and a
few GeV/s, with the maximum value reaching nearly 80~GeV/s.
\end{itemize}

Our results may be compared to those of previous works as follows. There is no
record in the literature of full kinematic evolution of thermal test particles
interacting with the shocked MHD plasma of a CWB with the objective of
understanding the acceleration process of high-energy particles. Estimates of
maximum energies and slopes, based on the first-order Fermi acceleration
process, as well as DSA, have been posed in the past \citep[e.g.,][and others,
see references therein]{EichlerUsov:1993, Dougherty_etal:2003, Reimer_etal:2006,
FalcetaGoncalvesAbraham:2012, Reitberger_etal:2014a, Reitberger_etal:2014b,
FalcetaGoncalves:2015, delPalacio_etal:2016, Pittard_etal:2020}.

Our findings of energies up to tens of TeV are in agreement with the upper limit
DSA estimates of \citet{FalcetaGoncalves:2015}, based on MHD simulations, as
well as to the approximate solutions of the transport equation, based on HD
simulations with assumed magnetization levels, given by
\citet{Reitberger_etal:2014a}. However, the latter have assumed much stronger
surface magnetic fields, in the order of 0.01T. Some discrepancies are expected,
given that our models contain self-consistent turbulent magnetic fields
generated in the shocks. With respect to the particle energy spectrum of the
protons, \citet{Reitberger_etal:2014a} obtained values in the range of -0.2 to
0.0, in good agreement with our models with stronger magnetic fields.

In the forthcoming studies, it is important to test the particle acceleration in
CWBs for particles other than proton. Therefore, in the first step, we plan to
determine the dependence of the maximum energy limit on the ratio $q/m$,
extending it toward heavier particles, such as carbon, as well as toward lighter
ones, like electrons. We expect that the processes responsible for energy losses
would be more important in the case of lighter particles, especially electrons.
Therefore, we plan to include some of the most important processes, such as
synchrotron, inverse Compton, and bremsstrahlung losses, directly in the
integration equations. In future works, it would be mandatory to perform these
models with a larger ensemble of particles in order to confirm the obtained
results, and more importantly to benchmark upper limit estimates, as well as
semi-analytic transport equation calculations for the simulation of nonthermal
emission from CWBs. Our results may also be directly compared to the
observations in the near future, for example, the Cherenkov Telescope Array
(CTA) \citep{Chernyakova_etal:2019}.

\section*{Data Availability Statement}

The raw data supporting the conclusions of this article will be made available
by the authors, without undue reservation.

\section*{Author Contributions}

D.F.G. formulated the scientific problem of particle acceleration in CWB. G.K.
prepared and performed numerical simulations of CWB model and the integration of
test particle trajectories. Both authors analyzed and interpreted the results
and contributed equally to the writing of the article.

\section*{Acknowledgments}
G.K. acknowledges support from the Brazilian National Council for Scientific and
Technological Development (CNPq no. 304891/2016-9) and FAPESP (grants
2013/10559-5 and 2019/03301-8). D.F.G. thanks the Brazilian agencies CNPq (no.
311128/2017-3) and FAPESP (no. 2013/10559-5) for their financial support.  This
work has made use of the computing facilities: the cluster of Núcleo de
Astrofísica Teórica, Universidade Cruzeiro do Sul, Brazil.
\bibliographystyle{frontiersinSCNS_ENG_HUMS}
\bibliography{ms}

\end{document}